\documentclass[twocolumn,pre,showpacs]{revtex4}
\usepackage{graphicx,color}
\usepackage{dcolumn}
\begin{document}
%
\title{
 Level spectroscopy of the square-lattice three-state Potts model with
 a ferromagnetic next-nearest-neighbor coupling
}
\author{
 Hiromi Otsuka, 
 Koutaro Mori,
 Yutaka Okabe, and 
 Kiyohide Nomura$^{\dagger}$
 }
\address{
 Department of Physics, Tokyo Metropolitan University, Tokyo 192-0397 Japan\\
 $^{\dagger}$Department of Physics, Kyushu University, Fukuoka 812-8581 Japan
 }
\date{\today}
\begin{abstract}
 We study the square-lattice three-state Potts model with the
 ferromagnetic next-nearest-neighbor coupling at finite temperature.
 Using the level-spectroscopy method, we numerically analyze the
 excitation spectrum of the transfer matrices and precisely determine
 the global phase diagram.
 Then we find that, contrary to a previous result based on the
 finite-size scaling, the massless region continues up to the decoupling
 point with ${\bf Z}_3\times{\bf Z}_3$ criticality in the
 antiferromagnetic region. 
 We also check the universal relations among excitation levels to
 provide the reliability of our result. 
\end{abstract}
\maketitle


 The concept of the Gaussian universality class provides a paradigm for the
 unified descriptions and understanding of the low-energy and
 long-distance properties of 
 one-dimensional (1D) quantum and
 two-dimensional (2D) classical systems.
 Its representation is given by the free-boson fixed point model, and its
 criticality by the conformal field theory (CFT) with the central charge
 $c=1$.
 The antiferromagnetic three-state Potts (AF3SP) model on the square
 lattice is one of those to exhibit the Gaussian criticality at zero
 temperature
 \cite{Lena67,Baxt70}.
 While the ground-state
 \cite{Lena67,Baxt70,Nijs82,Burt97,Delf01,Kola84}
 and finite-temperature properties
 \cite{Card01}
 were fully clarified, there still exists the considerable interest in its
 related extended-type models
 \cite{Nijs82,DelfNATO,Otsu04}.
 For instance, in
 \cite{Otsu04},
 the emergence of the ${\bf Z}_2$ and ${\bf Z}_3$ criticalities were
 studied by introducing the AF3SP model with a staggered polarization
 field. 

 In this paper we treat another model with a rather straightforward
 extension, i.e., the three-state Potts model with ferromagnetic 
 next-nearest-neighbor coupling defined on the square lattice $\Lambda$,
 whose reduced Hamiltonian
 ${\cal H}(K_1,K_2)=H/k_{\rm B}T$ is given as 
 \begin{equation}
  {\cal H}(K_1,K_2)
   = K_1 \sum_{\langle j,k\rangle} \delta_{\sigma_j,\sigma_{k}}
   -|K_2|\sum_{[       j,k      ]} \delta_{\sigma_j,\sigma_{k}}. 
   \label{eq_Hamil}
 \end{equation}
 The first and the second sums run over all nearest-neighbor and
 next-nearest-neighbor pairs, respectively, and the ternary
 variables $\sigma_j$ take the values 0,1,2 ($j\in\Lambda$).
 For the AF case ($K_1>0$), this model is thought to be in the same
 universality class as the ferromagnetic six-state clock (F6SC) model
 \cite{Card81}.
 Theoretical investigations including numerical ones have been
 performed to clarify the global phase diagram
 \cite{Nijs82,Gres81,Oliv84}. 
 However, its precise estimation is not available.
 This is mainly because two types of Berezinskii-Kosterlitz-Thouless
 (BKT) transitions take part in the phase diagram, and the numerical
 methods used so far were insufficient to treat the BKT transitions.
 For instance,
 the phenomenological renormalization-group (PRG) method has been
 frequently used for the determination of second-order transition
 points.
 However, as pointed out by several authors, it fails to estimate the
 BKT points
 \cite{Inou99}.
 Therefore, a new strategy should be employed.
 In investigations of 1D quantum systems, the level-spectroscopy
 method which takes logarithmic corrections into account has been used
 for precise estimates of the BKT points, and its possible application
 to the 2D classical system was also discussed
 \cite{Nomu95}.
 Therefore, we shall reconsider this long-standing model\ (\ref{eq_Hamil})
 with a new methodological strategy in order to clarify the global
 phase diagram.
 While the detail is given in the following,
 newly obtained data agree with the exact results available in some limits
 of model parameters, and
 show that the phase diagram is constructed by the crossover of the
 criticality from the  decoupling point $(K_1,K_2)=(0,\ln(1+\sqrt3))$
 with ${\bf Z}_3\times{\bf Z}_3$ criticality ($c=\frac85$), which is in
 contrast to the previous result based on the naive finite-size-scaling
 method
 \cite{Nijs82}.
 As a result, the precise determination of the phase diagram around the
 point turns out to be crucial to understand this model, whereas the
 Coulomb-gas picture provides a quantitative description only around
 the $K_1=+\infty$ line.

 Here we shall briefly refer to some relevant researches.
 First, the ground state of the AF3SP model ($K_1=+\infty$, $K_2=0$),
 which is equivalent to the ice-point six-vertex model and is thus
 critical
 \cite{Baxt70},
 becomes off-critical at finite temperature
 \cite{Nigh82}. 
 It was pointed out that there are two types of excitations controlled
 by the thermal scaling field $u=e^{-K_1}$, 
 i.e., the relevant and the marginal ones, and that both of these are
 necessary to explain the exotic corrections to scaling
 observed in the Monte Carlo data
 \cite{Card01}.
 The energy operator
 $\epsilon_j=\sum_{\langle j,k\rangle}\delta_{\sigma_j,\sigma_{k}}$
 (the scaling dimension $x_\epsilon=\frac32$)
 is the relevant excitation and brings about the correlation length
 $\xi\propto u^{-1/(2-x_\epsilon)}$ in its leading form.
 According to Delfino,
 the dual sine-Gordon Lagrangian density
 in term of the bosonic field and its dual field
 $\phi$, $\theta\in[0,2\pi/\sqrt2]$
 can describe this transition:
 \begin{equation}
  {\mathcal L}
   =
   \frac{1}{2\pi K}
   \left(\nabla \phi\right)^2
   +\frac{1}{2\pi\alpha^2}
   \left(y\cos6\sqrt2\phi+\bar{y}\cos\sqrt2\theta\right),
   \label{eq_L}
 \end{equation}
 where $y$ and $\bar{y}\propto u$ are the coupling constants
 and $\alpha$ is a short-distance cutoff
 \cite{Delf01,Card01}. 
 Since the Gaussian coupling $K$ is $\frac13$ for the AF3SP model at zero temperature,
 and thus the second term, the ${\bf Z}_6$ symmetry-breaking field, with the 
 scaling dimension 6 is highly irrelevant to this transition, we can
 drop it for the discussions of the AF3SP model. 
 However, it can become relevant for nonzero $K_2>0$ (see below). 

 Next, let us consider the case with $u=0$ ($K_1=+\infty$), where Eq.\
 (\ref{eq_Hamil}) is equivalent to the exactly solved $F$ model
 \cite{Lieb67}.
 Define $v=e^{K_2}$; then the $v$ dependence of the Gaussian
 coupling is given as $1/9K=1-(1/\pi)\cos^{-1}(v^2/2-1)$
 \cite{Baxt72}.
 Thus the point $v_2=2$ (i.e., $K=\frac19$) separates the massless
 ($v\le v_2$) region and the massive ($v>v_2$) region with sixfold
 degeneracy, and the latter is stabilized by the relevant term
 $y\cos6\sqrt2\phi$.
 Further, the nonlinear terms in Eq.\ (\ref{eq_L}) are both irrelevant
 in the region $v_1\le v\le v_2$ with
 $v_1=\sqrt{2+2\cos(5\pi/9)}\simeq1.2856$
 ($K=\frac14$), so the critical phase spreads at least for small $u$ and
 connects to the region in the limit $u\to0$
 \cite{Nijs82}.
 This argument is consistent with the so-called cell-spin analysis
 \cite{Park94}: 
 Park showed that Eq.\ (\ref{eq_Hamil}) with $K_2=K_1/2$ is
 approximately reduced to the F6SC model in terms of the cell spins representing
 the unit cells of the sixfold ground states.
 Therefore, as a member of the F6SC universality class
 \cite{Jose77},
 the present model\ (\ref{eq_Hamil}) is also expected to possess two
 BKT points $v_1(u)$ and $v_2(u)$, which is described by the
 dual sine-Gordon field theory Eq.\ (\ref{eq_L}).

 We shall first consider the system around $v_1(u)$ where
 $\cos6\sqrt2\phi$ is irrelevant, and thus it is described by 
 \begin{equation}
  {\mathcal L}_1
   =
   \frac{1}{2\pi K}
   \left(\nabla \phi\right)^2
   +\frac{\bar{y}}{2\pi\alpha^2}
   \cos\sqrt2\theta,~~~K\simeq\frac14. 
   \label{eq_L1}
 \end{equation}
 One of the authors (K.N.) investigated the properties of marginal
 operators on the BKT line: Especially, in this case, $(1/K)\left(\nabla
 \phi\right)^2$ 
 and 
 $\sqrt2\cos\sqrt2\theta$ (the bosonized expression of $\epsilon_j$)
 hybridize along the renormalization-group flow and
 result in two orthogonalized operators, i.e.,
 the ``${\mathcal M}$-like'' and the ``$\cos$-like'' operators
 \cite{Nomu95}. 
 Writing the former and the latter as $\bar{O}_0$ and $\bar{O}_1$,
 and defining the system on $\Lambda$ with $M$ $(\to\infty)$ rows of $L$
 sites wrapped on a cylinder, near the multicritical point, their
 renormalized scaling dimensions are given as 
 $\bar{x}_0(l)\simeq 2-y_0\left(1+\frac43 t\right)$
 and
 $\bar{x}_1(l)\simeq 2+y_0\left(2+\frac43 t\right)$,
 where
 $l=\ln L$, 
 $(y_0,y_1)=(1/2K-2,\bar{y})$, 
 and the small deviation from the BKT point
 $t=y_1/y_0-1$.
 Since these operators are parts of
 ${\mathcal L}_1$,
 the corresponding excitations possess the same symmetry with the ground
 state (see below).
 Another important operator is a relevant one:
 $O_2=\sqrt2\sin3\sqrt2\phi$
 whose dimension is expressed as 
 $x_2(l)\simeq\frac9{16}\left(2-y_0\right)$
 in the same region. 
 Consequently, the level-crossing condition
 $\bar{x}_0(l)=\frac{16}9x_2(l)$
 offers a finite-size estimate of the BKT point $v_1(u,L)$
 \cite{Nomu95}.

 Next we consider a region near $v_2(u)$ where
 $\cos\sqrt2\theta$
 is irrelevant.
 The effective Lagrangian is then given as
 \begin{equation}
  {\mathcal L}_2
   =
   \frac{1}{2\pi K}
   \left(\nabla \phi\right)^2
   +\frac{y}{2\pi\alpha^2}
   \cos6\sqrt2\phi,~~~K\simeq\frac19. 
   \label{eq_L2}
 \end{equation}
 In this case, the marginal operators
 $(1/K)\left(\nabla \phi\right)^2$
 and 
 $\sqrt2\cos6\sqrt2\phi$
 also hybridize and result in the ${\mathcal M}$-like
 and $\cos$-like operators,
 $O_0$ and $O_1$. 
 Then, their scaling dimensions near the transition point are
 $x_0(l)\simeq 2-y_0\left(1+\frac43 t\right)$
 and
 $x_1(l)\simeq 2+y_0\left(2+\frac43 t\right)$.
 Here,
 $(y_0,y_1)=(18K-2,y)$
 have been redefined.
 The operator $O_2$ has the dimension
 $x_2(l)\simeq \frac14\left[2-y_0\left(1+2 t\right)\right]$.
 Thus, the level-crossing condition for
 $v_2(u,L)$
 can be given by
 $x_0(l)=4x_2(l)$.
 Although there may exist some other level-crossing conditions for the
 determination of the BKT points, the levels focused on here and
 used in the following are relatively easy to be accessed by numerical
 calculations
 \cite{Nomu98}.

 Now, let us consider the system on $\Lambda$ with $M$ ($\to\infty$)
 rows of $L$ (an even number) sites, where $j\in\Lambda$ is specified by
 $l\in[1,L]$ and $m\in[1,\infty]$.
 For this system, we can define the transfer matrix ${\bf T}(L)$
 connecting 
 $(\sigma_{1,m},\cdots,\sigma_{L,m})$ 
 to
 $(\sigma_{1,m+1},\cdots,\sigma_{L,m+1})$,
 and denote its eigenvalues as
 $\{\lambda_n(L)\}$ or their logarithms as
 $\{E_n(L)=-\ln|\lambda_n(L)|\}$ ($n$ specifies a level).
 Then CFT provides direct expressions for $c$ and
 $x_n$ (the scaling dimensions) of critical systems as
 $E_{\rm g}(L)\simeq Lf-\pi v c/6L+b/L^3$
 and
 $\Delta E_n(L)\simeq \frac{2\pi v}{L}x_n$,
 where $E_{\rm g}$ is the smallest one corresponding to the ground state and
 $\Delta E_n(L)=E_n(L)-E_{\rm g}(L)$ is an excitation gap
 \cite{Card84,Blot86}.
 $1/v$, $f$, and $b$ are the geometric factor ($v=1$), the free energy per
 site, and a nonuniversal constant, respectively. 
 While we can calculate the renormalized scaling dimension $x_n(l)$
 from the excitation gap as
 $\Delta E_n(L)/(\frac{2\pi v }{L})$, 
 the symmetry properties are important for the specification of relevant
 excitations
 \cite{Otsu04}. 
 In addition to
 the translation ${\mathcal T}$
 and
 the space inversion ${\mathcal P}$, 
 the ${\bf S}_3$ symmetry associated with the global permutations of the
 ternary variables
 is also possessed by the present model\ (\ref{eq_Hamil}).
 By defining the clock variable as $s_j=e^{{\rm i} 2\pi \sigma_j/3}$, 
 the two generators of the {\bf S}$_3$ group are expressed as 
 the cyclic permutation
 ${\mathcal S}$:\ $s_j \mapsto e^{{\rm i} 2\pi/3} s_j$
 for the {\bf Z}$_3$ symmetry 
 and
 the charge conjugation
 ${\mathcal C}$:\ $s_j \mapsto s_j^*$
 for the charge conjugation symmetry ({\bf C}).
 These operations are then related to the changes of the phase
 variable as 
 ${\mathcal S}$:\ $\sqrt2\phi\mapsto\sqrt2\phi+2\pi/3$
 and 
 ${\mathcal C}$:\ $\sqrt2\phi\mapsto-\sqrt2\phi$
 \cite{Delf01}. 
 Apparently the marginal operators defined above are
 {\bf S}$_3$ invariant,
 and thus the corresponding excitation levels $\Delta E_n(L)$
 should be found in the subspace specified by this symmetry property. 
 On the other hand, the relevant operator $O_2$ is
 ${\bf Z}_3$ invariant, but is odd for the charge conjugation. 
 Further, since the sublattice symmetry property is determined by the
 ${\bf Z}_3$ charge
 \cite{Delf01},
 $O_2$ is odd for this symmetry operation.
 Therefore, we calculate the excitation levels according to the indices
 for these symmetry operations.

\begin{figure}
 \mbox{\includegraphics[width=2.8in]{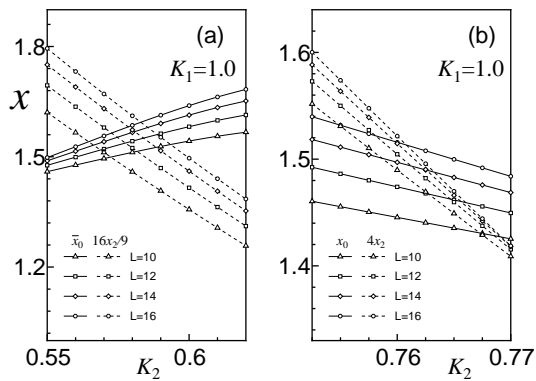}}
 \caption{
 Examples of the $K_2$ dependence of scaling dimensions
 $\Delta E_n(K_1,K_2,L)/(\frac{2\pi v}{L})$ at $K_1=1$.  
 (a) [(b)] shows $\bar{x}_0$ and $16x_2/9$ ($x_0$ and $4x_2$). 
 The correspondences between marks and system sizes are given in these
 panels.
 The crossing points give finite-size estimates $v_{1,2}(u,L)$.
 }
 \label{FIG1}
\end{figure}

 The exact diagonalization calculations of ${\bf T}(L)$
 are carried out for systems with $L=$8-16.
 In Figs.\ \ref{FIG1}(a) and \ref{FIG1}(b), we plot examples of the $K_2$
 dependencies of the scaled gaps
 $x_n=\Delta E_n(K_1,K_2,L)/(\frac{2\pi v}{L})$
 (or values multiplied by constants) at $K_1=1$.
 Then, we find the points [i.e., the finite-size estimates
 $v_{1,2}(u,L)$] at which one of the above conditions is satisfied.
 Here, it is worthy of note that, in addition to the logarithmic
 correction, there is another type of correction between a lattice system
 and a continuous model which stems from the $x=4$ irrelevant operator
 $L_{-2}\bar L_{-2}\bf{1}$ in terms of CFT
 \cite{Cardy1986NB}.
 Therefore, we shall extrapolate them to the thermodynamic limit
 according to the least-squares fitting of the polynomials in $1/L^2$.
 In Table\ \ref{TAB_I}, we give $v_{1,2}(u,L)$ at $K_1=64$ (i.e., a case
 almost equivalent to the $F$ model) in order to compare our data with
 the exact limiting values $v_{1,2}$.
 While considerable size dependencies are visible in the estimates, we can
 check that the extrapolated data do not deviate more than 0.7\% from the
 exact ones. 

\begin{figure}
 \includegraphics[width=2.8in]{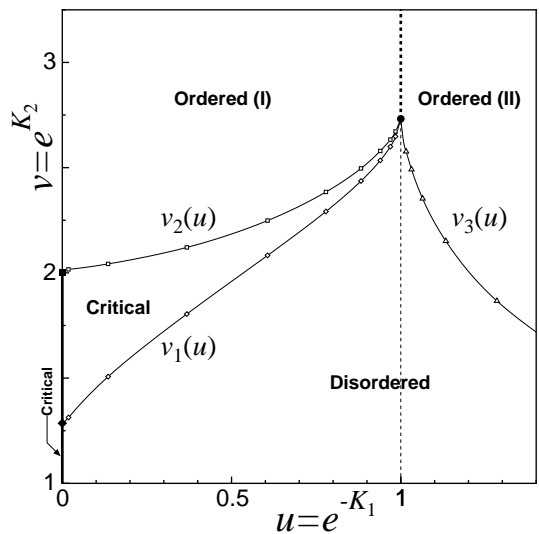}
 \caption{
 The phase diagram of the model\ (\ref{eq_Hamil}).
 The open diamonds (squares) with a curve exhibit the BKT
 line $v_1(u)$ $[v_2(u)]$. 
 The filled diamond and the square on the $v$ axis correspond to
 $(0,v_1)$ and $(0,v_2)$, respectively.
 The filled circle at $(1,1+\sqrt{3})$ denotes the  decoupling point
 with ${\bf Z}_3\times{\bf Z}_3$ criticality.
 The open triangles with a curve exhibits the second-order transition
 line $v_3(u)$.
 The thick dotted line shows the first-order phase transition boundary.
 }
 \label{FIG2}
\end{figure}

 In Fig.\ \ref{FIG2}, we give the phase diagram of our model\
 (\ref{eq_Hamil}) in the space $(u,v)=(e^{-K_1},e^{K_2})$.
 The open diamonds (squares) with a curve exhibit the phase boundary
 $v_1(u)$ $[v_2(u)]$, which separates the critical region from the
 disordered phase [from the ordered phase ``Ordered (I)'' with sixfold
 degeneracy]. 
 The filled diamond and the square on the $v$ axis show $(0,v_1)$ and
 $(0,v_2)$, respectively.
 The filled circle at $(u,v)=(1,1+\sqrt{3})$ denotes the decoupling
 point with {\bf Z}$_3\times${\bf Z}$_3$ criticality
 \cite{Pott52}.
 To complete the phase diagram,
 we also calculate the second-order phase transition point $v_3(u)$
 with the ferromagnetic three-state Potts (F3SP) criticality in the
 ferromagnetic region $u>1$ (i.e., triangles) by employing the PRG
 method
 \cite{Nijs82}.
 The ordered phase ``Ordered (II)'' has threefold degeneracy, and the
 thick dotted line shows the first-order phase transition boundary. 
%
\begin{table}
 \caption{
 The finite-size estimates of the BKT points for systems close
 to the $F$ model $[v_{1,2}(0,L)]$.
 Extrapolations to $L\to\infty$ are performed according to the
 least-squares fitting of the polynomials in $1/L^2$.
 } 
 \begin{tabular}{ccccccc}
  \hline\hline
    $L$&8&10&12&14&16&  $\infty$ \\ 
  \tableline
  $v_1(0,L)$ &1.31465&1.30449&1.29871&1.29517&1.29287&1.285\\
  $v_2(0,L)$ &2.02507&2.02626&2.02541&2.02404&2.02262&2.013\\
  \hline\hline
 \end{tabular}
 \label{TAB_I}
\end{table}
%
 From this figure we can find the following.
 While the cusplike behavior of the phase diagram is observed around
 the decoupling point
 \cite{Leeu75},
 two BKT lines $v_{1,2}(u)$ do not merge into a single curve
 contrary to the observation in the previous work based on the PRG
 calculation
 \cite{Nijs82}.
 Therefore, we conclude that the massless critical region continues up
 to the decoupling point and that neither a critical end point nor a
 first-order transition line exists in the antiferromagnetic region
 $u<1$. 
 Further, contrary to the previous calculation,
 the $u$ dependence of $v_2(u)$ is monotonic and does not show a
 minimum around $u\simeq0.3$.
 As a result, the phase diagram obtained by the
 level-spectroscopy method is considerably different from the previous
 one, and shows that the Gaussian critical region is stabilized in a
 wider area of the parameter space.

 Before moving to discussion, we shall check the consistencies among
 excitation levels on the BKT points. 
 Since the amplitudes of logarithmic corrections are given by the
 operator-product-expansion coefficients, some universal relations among
 the scaling dimensions have been discovered
 \cite{Nomu95,Zima87}:
 For instance,
 $[2\bar{x}_0(l)+\bar{x}_1(l)]/3\simeq2$ at $v_1(u)$
 and 
 $[3x_2(l)+x_3(l)]/4\simeq\frac12$ at $v_2(u)$. 
 Here $x_3(l)$ is the dimension of
 $O_3=\sqrt2\cos3\sqrt2\phi$ 
 [the bosonized expression of the staggered polarization
 $P_j=(-1)^j\sum_{[j,k]}\delta_{\sigma_j,\sigma_{k}}$]
 which is ${\bf S}_3$ invariant and odd for the sublattice symmetry 
 \cite{Delf01,Note01}.
 In Table\ \ref{TAB_II}, we give the scaling dimensions at $K_1=1$. 
 Although $\bar{x}_0(l)$ and $\bar{x}_1(l)$ [$x_2(l)$ and $x_3(l)$]
 considerably deviate from the value for the free-boson case 2 ($\frac12$)
 due to the logarithmic corrections, their main parts cancel each other,
 so the average 
 $\bar{x}_{\rm av}=(2\bar{x}_0+\bar{x}_1)/3$ 
 $[    x_{ \rm av}=(3x_2+x_3)/4]$
 takes a value close to 2 ($\frac12$). 
 The consistency checks can be satisfied only if the systems are
 on the BKT lines, and the investigated levels based on symmetry
 properties have the expected physical interpretations. 
 Therefore, Table\ \ref{TAB_II} is helpful to demonstrate the reliability
 of our phase diagram.
%
\begin{table}[t]
 \caption{
 Examples of the $L$ dependences of the scaling dimensions $\bar{x}_{0}$
 and $x_{2}$ and the averages $\bar{x}_{\rm av}$ and $x_{\rm av}$ (see
 text) on the BKT points $v_{1,2}(u)$ ($K_1=1$).
 We extrapolate the finite-size estimates to $L\to\infty$ using the
 least-squares fitting of the polynomial in $1/L^2$.
 } 
 \begin{tabular}{rcccccc}
  \hline\hline
  $L$&8&10&12&14&16&$\infty$\\ 
  \tableline
  $\bar{x}_{     0}(l)$ &1.48366&1.52857&1.56357&1.59141&1.61404& --- \\
  $\bar{x}_{\rm av}(l)$ &1.78633&1.83948&1.87883&1.90716&1.92759&2.006\\
  \tableline
  $    {x}_{     2}(l)$ &0.38137&0.38913&0.39454&0.39851&0.40155& --- \\
  $    {x}_{\rm av}(l)$ &0.50271&0.50078&0.49933&0.49811&0.49704&0.492\\
  \hline\hline
 \end{tabular}
 \label{TAB_II}
\end{table}

 Finally, we discuss the nature of the phase diagram.
 As the real-space RG calculations predicted
 \cite{Oliv84,Leeu75},
 the decoupling point $(1,1+\sqrt3)$ with {\bf Z}$_3\times${\bf Z}$_3$
 criticality (i.e., $c=\frac85$) becomes unstable against relevant
 competing perturbations and exhibits crossovers to the behaviors
 controlled by the critical fixed points with lower symmetries
 \cite{Zamo86}.
 One of those perturbations is the energy operator with scaling
 dimension $\frac45$.
 For the ferromagnetic case $u>1$, another one may be a product of
 order parameters on each sublattice which has the dimension
 $\frac{4}{15}$. 
 Therefore, the crossover exponent
 $\phi_{\rm F}=(2-\frac{4}{15})/(2-\frac45)=\frac{13}{9}$ 
 can predict the shape of the boundary around the point, 
 and
 $v_3(u)$ agrees with this
 \cite{Nijs82}. 
 Further, we obtain, for instance, $c\simeq 0.800$ at $v_3(u)$
 ($K_1=-0.25$), so the boundary can be regarded as a massless RG flow to
 the F3SP fixed point with $c=\frac45$.
 On the other hand, we also analyze $v_{1,2}(u)$, and similar power law
 behaviors can be recognized. 
 Although the estimated exponents take close values to $\frac{13}{9}$,
 the linearities of the log-log plots are rather lower than in the case
 of $v_{3}(u)$, and thus the estimates may acquire errors up to
 10\%. 
 This may be due to the existence of the marginal operator in the
 intermediate critical region, which is absent in the ferromagnetic case. 
 Since the central charge takes the values, e.g., $c\simeq 1.007$ and
 1.009 at $v_1(u)$ and $v_2(u)$ ($K_1=1$), respectively, the critical
 region can also be seen as a realization of the crossover to the
 Gaussian fixed point with $c=1$
 \cite{Zima87,Kita99}.
 However, its precise numerical analysis is still difficult, so this
 issue will be left as a future problem.

 In summary, we investigated the square-lattice three-state Potts
 model with the ferromagnetic next-nearest-neighbor coupling. 
 The phase diagram obtained here reveals new features that could not be
 found in previous research.
 With the use of the level-spectroscopy method, we can break new ground
 for studying 2D classical systems with complex interactions.
 For more detailed study, we are now performing the Monte Carlo
 simulations, and we obtain preliminary data consistent with the present
 result of the phase diagram; they will be given in a forthcoming
 article.

 One of the authors (H.O.) thanks 
 M. Nakamura
 for stimulating discussions. 
 This work was supported by Grants-in-Aid from
 the Ministry of Education, Culture, Sports, Science and Technology of
 Japan. 



\begin{references}
 \newcommand{\REF }[4]{#1 {\bf #2}, #3 (#4)}
 \newcommand{\PRL }[3]{\REF{Phys. Rev. Lett.\   }{#1}{#2}{#3}}
 \newcommand{\PRA }[3]{\REF{Phys. Rev.\        A}{#1}{#2}{#3}}
 \newcommand{\PRB }[3]{\REF{Phys. Rev.\        B}{#1}{#2}{#3}}
 \newcommand{\PRD }[3]{\REF{Phys. Rev.\        D}{#1}{#2}{#3}}
 \newcommand{\PRE }[3]{\REF{Phys. Rev.\        E}{#1}{#2}{#3}}
 \newcommand{\PLA }[3]{\REF{Phys. Lett.\       A}{#1}{#2}{#3}}
 \newcommand{\PLB }[3]{\REF{Phys. Lett.\       B}{#1}{#2}{#3}}
 \newcommand{\NPB }[3]{\REF{Nucl. Phys.\       B}{#1}{#2}{#3}}
 \newcommand{\ZPB }[3]{\REF{Z.    Phys.\       B}{#1}{#2}{#3}}
 \newcommand{\CMP }[3]{\REF{Comm. Math. Phys.\  }{#1}{#2}{#3}}
 \newcommand{\JMP }[3]{\REF{J. Math. Phys.\     }{#1}{#2}{#3}}
 \newcommand{\JPSJ}[3]{\REF{J. Phys. Soc. Jpn.\ }{#1}{#2}{#3}}
 \newcommand{\JSP }[3]{\REF{J. Stat. Phys.\     }{#1}{#2}{#3}}
 \newcommand{\JPA }[3]{\REF{J. Phys.\          A}{#1}{#2}{#3}}
 \newcommand{\JPC }[3]{\REF{J. Phys.\          C}{#1}{#2}{#3}}
 \newcommand{\IBID}[3]{\REF{{\it ibid.}}{#1}{#2}{#3}}
 \newcommand{\etal}{{\it et al.}}

 \bibitem{Lena67}
 E.H. Lieb,
 \REF{Phys. Rev.}{162}{162}{1967};
 E.H. Lieb, 
 \PRL{18}{692}{1967}. 

 \bibitem{Baxt70}
 R.J. Baxter,
 \REF{J. Math. Phys.}{11}{3116}{1970};
 R.J. Baxter,
 \REF{Proc. R. Soc. London, Ser. A}{383}{43}{1982}.

 \bibitem{Nijs82}
 M.P.M. den Nijs, M.P. Nightingale, and M. Schick, 
 \PRB{26}{2490}{1982}.

 \bibitem{Burt97}
 J.K. Burton, Jr., and C.L. Henley,
 \JPA{30}{8385}{1997}.

 \bibitem{Delf01}
 G. Delfino, 
 \JPA{34}{L311}{2001}. 

 \bibitem{Kola84}
 J. Kolafa,
 \JPA{17}{L777}{1984};
 J.-S. Wang, R.H. Swendsen, and R. Koteck\'y, 
 \PRL{63}{109}{1989}. 

 \bibitem{Card01}
 J.L. Cardy, J.L. Jacobsen, and A. Sokal,
 \JSP{105}{25}{2001},
 and the references therein.

 \bibitem{DelfNATO}
 G. Delfino, in 
 {\em Proceedings of the NATO Advanced Research Workshop on Statistical
 Field Theories},
 edited by A. Cappelli \etal~(Kluwer, Dordrecht, 2002), eprint
 hep-th/0110181. 

 \bibitem{Otsu04}
 H. Otsuka and Y. Okabe,
 \PRL{93}{120601}{2004}. 

 \bibitem{Card81}
 J.L. Cardy,
 \PRB{24}{5128}{1981}.

 \bibitem{Gres81}
 G.S. Grest and J.R. Banavar,
 \PRL{46}{1458}{1981};
 I. Ono, 
 \JPSJ{53}{4102}{1984}. 

 \bibitem{Oliv84}
 P.M. Oliveira, C. Tsallis, and G. Schwachheim 
 \PRB{29}{2755}{1984}.

 \bibitem{Inou99}
 J.C. Bonner and G. M\"uller,
 \PRB{29}{5216}{1984};
 J. S\'olyom and T.A.L. Ziman,
 \IBID{30}{3980}{1984};
 H. Inoue and K. Nomura, 
 \PLA{262}{96}{1999}.

 \bibitem{Nomu95}
 K. Nomura, 
 \JPA{28}{5451}{1995}.

 \bibitem{Nigh82}
 M.P. Nightingale and M. Schick,
 \JPA{15}{L39}{1982};
 C. Jayaprakash and J. Tobochnik,
 \PRB{25}{4890}{1982}.

 \bibitem{Lieb67}
 E.H. Lieb, 
 \PRL{18}{1046}{1967}.

 \bibitem{Baxt72}
 R.J. Baxter,
 \REF{Ann. Phys. (N.Y.)}{70}{193}{1972}. 

 \bibitem{Park94}
 H. Park, 
 \PRB{49}{12881}{1994}.

 \bibitem{Jose77}
 J.V. Jos\'e, L.P. Kadanoff, S. Kirkpatrick, and D.R. Nelson, 
 \PRB{16}{1217}{1977}.

 \bibitem{Nomu98}
 For the possibility of more efficient evaluation, see  
 K. Nomura and A. Kitazawa, 
 \JPA{32}{7341}{1998};
 H. Matsuo and K. Nomura
 (unpublished).
 
 \bibitem{Card84}
 J.L. Cardy,
 \JPA{17}{L385}{1984}.

 \bibitem{Blot86}
 H.W.J. Bl\"ote, J.L. Cardy, and M.P. Nightingale,
 \PRL{56}{742}{1986};
 I. Affleck,
 \IBID{56}{746}{1986}.

 \bibitem{Cardy1986NB}
 J.L. Cardy, 
 \NPB{270~{\rm[FS16]}}{186}{1986}.

 \bibitem{Pott52}
 R.B. Potts, 
 \REF{Proc. Cambridge. Philos. Soc.}{48}{106}{1952}; 
 F.Y. Wu, 
 \REF{Rev. Mod. Phys.}{54}{235}{1982}. 

 \bibitem{Leeu75}
 J.M.J. van Leeuwen, 
 \PRL{34}{1056}{1975}.

 \bibitem{Zima87}
 T. Ziman and H.J. Schulz, 
 \PRL{59}{140}{1987}.

 \bibitem{Note01}
 By denoting the interpenetrating sublattices of $\Lambda$ as
 $\Lambda_\pm$, $(-1)^j$ is defined to take 
 $+1$ for $j\in\Lambda_+$
 and
 $-1$ for $j\in\Lambda_-$. 

 \bibitem{Zamo86} 
 A.B. Zamolodchikov,
 \REF{Pis'ma Zh. Eksp. Teor. Fiz.}{43}{565}{1986}
 [\REF{JETP Lett.}{43}{730}{1986}].

 \bibitem{Kita99}
 A crossover from $c=\frac32$ to 1 was discussed by
 A. Kitazawa and K. Nomura, 
 \PRB{59}{11358}{1999}.
\end{references}
\end{document}